\documentclass[aps,prl,twocolumn,10 pt,superscriptaddress,showpacs]{revtex4}
\usepackage{amssymb}
\usepackage{graphicx}

\begin{document}

\title{An experimental study of narrow band noise due to the moving vortex lattice in Niobium.}
\author{Alain Pautrat}
\affiliation{Laboratoire CRISMAT, UMR 6508 du CNRS, ENSICAEN et Universit\'e de Caen, 6 Bd Mar\'echal Juin,14050 Caen 4, France.} 
\author{Joseph Scola}
\affiliation{Groupe d'Etude de la Mati\`ere Condens\'ee (GEMAC), CNRS-Universit\'e de Versailles St.Quentin, 45 Avenue des Etats-Unis, 78035 Versailles, France.}

\begin{abstract}

We report measurements of voltage noise due to vortex motion in Niobium, a conventional low-T$_c$ superconductor.
A coherent oscillation leading to narrow band noise (NBN) is evidenced. Its characteristic frequency is a linear function of the overcritical transport current in the flux-flow regime, and hence scales as the main
 velocity of the vortex flow. The associated length scale is not the intervortex distance but the width of the sample,
 indicating temporal coherence at a large scale. NBN is also observed in the non linear part of the $V(I)$ at the onset
 of depinning, in apparent disagreement with a stochastic creep motion of flux bundles.
NBN exists in the peak effect region, showing that long range temporal correlations are preserved in this regime.

\end{abstract}

\newpage
\maketitle

\section{Introduction}
 The properties of the vortex lattice (VL) have always attracted a lot of attention. Among many interesting aspects,
 similarities with other periodic systems as charge density waves and Wigner crystals, or even colloids on surfaces has been recently emphasized.
 Therefore, VL has been used as a model system for studying disordered elastic systems and collective dynamics \cite{giam}.
 One of the challenges is to understand the response of a system, submitted to quenched disorder, to an external force.
Complex phase diagrams have been theoretically predicted, which differ in details,
 but tend to an average agreement: the pinning induces a defective structure which reorders when 
submitted to a large applied force \cite{kv,gld,balents}.
 The predictions which can be checked experimentally are concerning the variation of the VL velocity (its main value or the fluctuations of its value),
 or the variation of the structural order (the structure factor in a diffraction experiment) as function of the driving force.
 In the simplest case, the driving force can be assumed
 to be proportional to the applied current, what is relevant if all the current flows homogeneously through the bulk.
 Note however that this neglects the important cases of edge or surface pinning \cite{zeldov,nous},
 where most of the (critical) current is confined close to the boundaries of the sample.
 In conventional and homogeneous soft superconductors,
 the $V(I)$ curve is a line shifted from the origin by the critical current $I_c$.
 Two regimes (non dissipative: $V=0$ for $I<I_c$, and flux-flow: $V=R_{ff}(I-I_c)$ for $I>I_c$) describe
 such curves. When the current is lower than $I_c$, the VL is pinned, the main velocity and its fluctuations are zero.
 At higher current, in the flux-flow regime, the vortices tend to form a long range ordered state,
 as experimentally confirmed by the existence of orientational and positional long range order from neutron scattering experiments \cite{nous2}.
 The exact nature of the ordered state
 has been largely debated. Theories predict a crystalline \cite {kv} or a smectic order \cite{balents},
 what refer closely to elastically coupled or decoupled
 channels in a moving glass \cite{gld}. Most of the $V(I)$ curves 
present a third regime close to the threshold were the voltage is a non linear function of $(I-I_c)$ and
 is usually associated to a creep motion of flux bundles.
  Since the existence and the shape of this non linear part is found strongly sample dependent, the role
 of homogeneity of the critical current 
 at the sample scale can be important \cite{jones,nous}. In addition to the small shear modulus of the VL, this causes
 a depinning of large slices of vortices and the global non linear response can be explained by a sum of small linear regions.
  In the case of strong disorder and moderate velocity (close to the depinning threshold),
 a large number of dislocations can lead to plasticity. Simulations have found that
 it corresponds to a peak in $d^2V/dI^2$,
 or equivalently to a convex upward $V(I)$ curve \cite{olson,ryu}.
An anomalously large amount of broad band noise (BBN) is also expected \cite{olson,kolton2,Enrik}. These BBN signatures
 are observed in a part of the phase diagram where the peak effect in the critical current occurs.
 $NbSe_2$ (pure and $Fe$ doped) is certainly the most studied material, and series of quite thorough experimental investigations have been performed.
 They converge to the idea that the anomalous transport properties are dominated by the kinetics between two
 macroscopic states with different pinning strengths \cite{shobo}. Both the large non-gaussian
 noise and the peak in the differential resistance have been explained by a mechanism of "edge contamination"  \cite{paltiel}. 
 
Since the nature of disorder is one of the central point, an interesting and sensitive experimental probe
 is the measure of the noise generated during the vortex motion. As it is deduced from the relation 
\begin{equation}
\label{josephsonequation}
E = -v_L \times B
\end{equation}
 any fluctuation of vortex density or velocity
 will be the cause of voltage noise \cite{clem,bernard2,jojo}. Noise due to VL motion can exist in the form
 of a BBN and/or of a narrow band noise (NBN).
BBN is quite general and possibly ambiguous: detailed investigation is needed to elucidate its origin \cite{clem,bernard2}.
In contrast, the observation of NBN is itself an information. This is a probe of temporal coherence. The washboard noise
 whose frequency is $f=v_L/\lambda $ ($\lambda $ is the wavelength, i.e. the intervortex distance $a_0$ for the VL and $v_L$ is the flow velocity)
 is one example of NBN.
 Such a noisy signature has been largely studied in the case of charges density waves sliding \cite{cdw}.
The study of the noise in moving VL has a long story and BBN has been largely reported. However, only few experiments have reported the existence of NBN.
One experiment has associated the NBN observed in the High $T_c$ superconductor $Bi-2212$ to a washboard effect \cite{maeda}.
 Taking $a_0$ as the characteristic scale, the associated velocity
 was found very low and strongly field dependent. It was concluded that the VL should move
 in a Creep regime, but where washboard noise is $\textit{a priori}$ not expected.
 The second experiment was made in a very peculiar experimental geometry
 in a thin film of YBCO with the field along the large surfaces \cite{danna}. The (here asymmetric due to the presence of a substrate)
 role of the surfaces for the NBN generation was pointed out.
Cross-correlation functions of flux-flow noise
 was measured some years ago by Heiden et al \cite{nbnheiden}. They used movable contact
 pairs to measure time of flight effects. Most of the samples do not report any effect
 (polycrystalline Niobium and Vanadium foils), except a monocrystalline Niobium
 which shows a coherence peak likely associated to a time of flight between
 the two contacts pairs which were close to the sample boundaries.

 To extract a scale from a flight effect as in \cite{nbnheiden}, it is necessary to know the velocity. For VL, this latter is clearly defined
 in the flux-flow regime from Eq.~\ref{josephsonequation}, $v_L=E/B$. Along a $V(I)$ curve at fixed $T$ and $B$, 
the velocity can then be directly controlled by changing the applied current.
 Here, we report direct observations of NBN in Niobium.
It is observed in addition to the more usual BBN which has been analyzed with some details before \cite{jo2}.
Interestingly, the BBN is observed in the presence of the peak effect in the critical current.

 \section{Experimental}

The samples used were strips of Niobium ($T_c=9.2K$, RRR $\approx$ 800).
 They have been characterized by specific heat, magnetization and electronic transport. Samples are polycrystalline.
The grain size is a few micrometers as revealed by scanning electron microscope images.
Our samples were mechanically then chemically polished in order to limit surface pinning and thereby having the critical current in a measurable range.
The critical fields delimiting mixed state and
 surface superconductivity phases were
 respectively $B_{c2}=0.290 \pm 0.005 T$ and $B_{c3}\approx 0.52 T$ at $4.2K$ \cite{jo1}.
 These values are those of pure Niobium.
 In mixed and surface superconductivity states, non-ohmicity and broad band
 voltage noise due to surface pinning currents have been observed and reported elsewhere \cite{jo1,jo2}.
Samples with different widths were investigated. The sample $\sharp 1$ and  $\sharp 2$
 have widths of W= $0.25$ mm and $1.25$ mm, respectively, and length$\times$thickness=
 $5$ $\times$ $0.22$ mm$^2$. The sample geometry is shown in fig.1. The sample $\sharp 1$ with the small
 width exhibits the peak effect 
in the critical current. To ensure that the applied current ($I=0-10 A$) is free from noise,
 it was delivered by car batteries ($12V-70 Ah$) in series with thermalized power resistance.
The measured voltage was amplified by ultra low noise preamplifier (NF Electronics Instruments,
 model $SA-400 F3$) enclosed in a thick screening box.  
Signals were recorded with a $PCI-4551$ (National instrument) analog input channel, anti-aliased
 and Fourier transformed in real time.
The quantity of interest is here the autopower spectra $S_{vv}(f)$ ($V^2/Hz$) of the voltage noise
 measured along the direction of the applied current (longitudinal noise). Contact noise and sample noise were systematically
 checked at large current in the normal state ($B>B_{c3}$) and in the Meissner state ($B<B_{c1}\approx 0.14 T$) and were found negligible compared to background noise.

\section{Results}
\subsubsection{Phenomenological description of the NBN}

Fig.2 shows $S_{vv}(f)$ for $I= 0.45, 0.48$ and $0.53 A$ and $B=0.28 T$ in the mixed state (sample $\sharp 1$).
 A background noise measured under similar biasing conditions
 but at high field ($B>B_{c3}$) has been subtracted.
$S_{vv}(f)$ clearly exhibits two components.
The first one is a BBN corresponding to the flux-flow noise and has been deeply investigated by several authors \cite{clem,jo2,bernard2}.
The second feature of the voltage spectra is characterized by a sharp peaks centered at frequencies ranging from $10 Hz$ to about $40 Hz$.
These narrow-band contributions to the noise spectra, or NBN, appears simultaneously at frequencies which are the multiples of a fundamental frequency $f_c$.
First and sometimes second harmonics can be observed. The analysis of a time series (from 30 minutes long acquisitions) of spectra indicates that the NBN is stationary and gaussian (we observe frequency independent second spectra, \cite{secondspectre}).

 NBN has been measured for different biasing currents along the $V(I)$ curves.
A typical $V(I)$ and the corresponding $f_c(I)$ are shown in fig.3.
 In the flux-flow regime, we find that $f_c$ is proportional to $(I-I_c)$ (Fig.4), hence to the main VL velocity $v_L$ given by:
\begin{equation}
\label{vortexvelocity}
v_L=\frac{E}{B}=\frac{R_{ff}(I-I_c)}{B\cdot d},
\end{equation}
where $d$ is the distance between voltage contacts.
The slope $v_L/f_c$ gives a value having the dimension of a distance. For $B<B_{c2}$, we find $f_c/v_L=1.30 \pm 0.05$ mm.
This agrees extremely well with the width of the sample $W=1.25$ mm for the sample $\sharp 2$. 
We obtain also a good agreement with sample $\sharp 1$ which is about five times narrower.
 The linear fit gives $f_c/v_L=0.35 \pm 0.05$ mm to compare with $W=0.25$ mm.
A NBN contribution could be linked with moving inhomogeneities coherently fluctuating over a pinning length scale.
 The fact that this length scale is found to be the width of the sample could indicate that the
 inhomogeneities are forming at the foil edges where the lattice begins to form.
 We suspect that small irregularities of the edges could be the cause of the fluctuations. 

In the following, we will focus on the qualitative properties of the the narrow band contribution
 to the noise spectra.

\subsubsection{Discussion}

Let us compare the results with usual descriptions of NBN.
NBN is often found to replace BBN for clean or ordered VL indicating that the motion becomes coherent \cite{olson}.
The present results clearly show that both kinds of noise can be simultaneously observed.
Besides, the characteristic length associated to the phenomenon referred to as NBN in simulations is the inter-vortex distance.
 This is five orders of magnitude smaller than the one observed here.
 In our experimental conditions, this washboard effect is expected at much higher frequency than the low frequency noise currently measured
 and its measurement requires a high-frequency dedicated set-up.
Hence, it must be noticed that we are dealing with a different phenomenon, but in any case
the NBN evidences the coherence of the VL at the scale of the sample.
Its simultaneous occurrence with BBN demonstrates that the BBN does not witnesses disorder.
Indeed, the BBN persists at high VL velocity in a well ordered state.
This has been consistently explained by description of the BBN relying
 on ultra-fast surface fluctuations of the critical current reflecting the pinning
 potential independently of the line velocity and of macroscopic reordering \cite{bernard2,jo2}.
 We do not find any direct correlation between the NBN and the BBN magnitudes when the current is increased.
After surface irradiation of the sample, we observed a dramatic change of the BBN properties,
 with a strong increase of its magnitude \cite{jo2}, and the NBN is not observed anymore.
 It may indicate that a strong surface disorder breaks the long range temporal coherence needed for the NBN,
 but this deserves more systematic investigations.

NBN is also observed close to the depinning current, in the non-linear part of the $V(I)$ curve before reaching the flux-flow (fig.3).
Its characteristic frequency $f_c$ is observed higher than expected from extrapolating the flux-flow behavior to low velocity.
This discrepancy is not surprising because the VL dynamical regime at low current is expected to differ from that in flux-flow.
All models of vortex motion agree that the velocity should be inhomogeneous in this regime.
As discussed above, this regime has been described either as a creep motion, as a plastic flow or as a progressive depinning of a succession
 of large slices due to a non-uniform distribution of critical current over the length of the sample.
The observation of NBN implies that the velocity of large parts of the VL fluctuates coherently and it is difficult to reconcile with
 the either creep regime or plastic flow.
 On the contrary, moving and still parts of the VL are expected to coexist, so that the mean velocity defined by Eq. \ref{vortexvelocity}
 should under-estimate the actual vortex velocity (as $d$ over-estimates the size of the moving VL),
 thus explaining the observed behavior. In addition, to the extend that the width of the sample is always the pertinent scale,
 the quality factor is $Q=f_c/\Delta f_c\approx v_L/\Delta v_L$. It is always resolution limited with $\Delta f_c\approx 4Hz$.
 This shows that $\Delta v_L/v_L\leq 0.1$ and that the velocity dispersion is not much larger than in flux-flow. 
 We note that washboard oscillation was previously observed by fast imaging of VLs by STM in the so-called creep regime at a very local scale \cite{stm}.
Due to the measuring technique,  it was not possible to know if coherence is preserved at the sample scale.
Maeda \emph{et al.} \cite{maeda} have observed NBN in the voltage noise in Bi-2212
 at very low field ($B\approx 20-80 G$
 corresponds to $a_0 \approx 1-0.5 \mu m$).
 To associate this NBN to the washboard frequency,
a very low velocity of the VL and a strong field dependence should be assumed
 (as low as $10^{-4} m/s$), which is more consistent with a creep regime.
 They conclude that temporal order is observed in a regime where a
 stochastic motion is though to take place.
Finally, all these different experiments show that the "Creep" regime is much more coherent than it is usually thought.
 Note neutron scattering experiments in conventional low-T$_c$ materials indicate also that the VL is much more
 regular than expected for a model of independent flux bundles motion at the beginning of its motion \cite{thorel,nous2,gougou}.

\subsubsection{Scarcity of the NBN in VL studies}
Only few occurrences of NBN in moving VLs have been reported in the literature.
If the frequency of NBN is $f_c=v_L/W$, this means that a compromise between the VL velocity and
 the sample width has to be found for allowing its observation.
Noise experiments, adapted to VLs studies, allow to measure noise spectrum typically for frequencies $f<1-10 KHz$.
Low frequency will be obtained with a large width and a low VL velocity.
A large width implies a large critical current which is bad for noise experiments, because of heating problems. 
 A low VL velocity can be obtained in samples with a low normal state resistivity ($v_L=E/B=\rho_{ff}(J-J_c)/B \propto \rho _n$),
 i.e. pure metals rather than alloys. This is why Niobium is a good candidate for the observation of such an effect.
On the other hand, a low resistivity generally favors a low noise amount and makes the measurements challenging \cite{notebene}.
The experimental window is then very small.

\subsubsection{NBN and peak effect}
  
Interestingly, the sample with the smallest width exhibits a peak effect in the critical current.
 This is shown in fig.5. The peak
 effect in our Niobium sample has several difference from the one reported in $NbSe_2$.
 In $NbSe_2$, an extremely large noise amount is observed in the lower part of the peak effect,
 where a dynamic phase coexistence occurs. In addition, this noise has a non-gaussian character \cite{shobo2,mikeNb}.
 Despite a systematic analysis of 
 effects by measurements of second spectra \cite{secondspectre}, we do not observed such effects here.
Moreover, we do not see a change in the $V(I)$ shape when crossing
 the peak, contrary to the observation made in $NbSe_2$. In our Niobium sample, we observe a peak effect with conventional transport properties,
 including the presence of NBN which proves a coherent VL motion. This confirms that the anomalous transport properties which
 can be observed close to a peak effect are directly coming from
 the phase coexistence, which appears to be prominent in some samples but almost absent in others, for reasons that deserve more attention.

 \section{Conclusion}

 In conclusion, we have reported measurements of Narrow Band Noise (NBN) in a low $T_c$ superconductor.
 This NBN characteristic scale is the width of the sample. The presence of harmonics confirms a temporal coherence.
 This is also observed in a sample which exhibits
 the peak effect in the critical current. As a consequence, the moving state of vortices in the peak effect can not be described by an incoherent
 disordered state, at least in the case of the moderate peak effect of Niobium.

\newpage

\begin{figure}[!t]
\begin{center}
\includegraphics*[width=9.0cm]{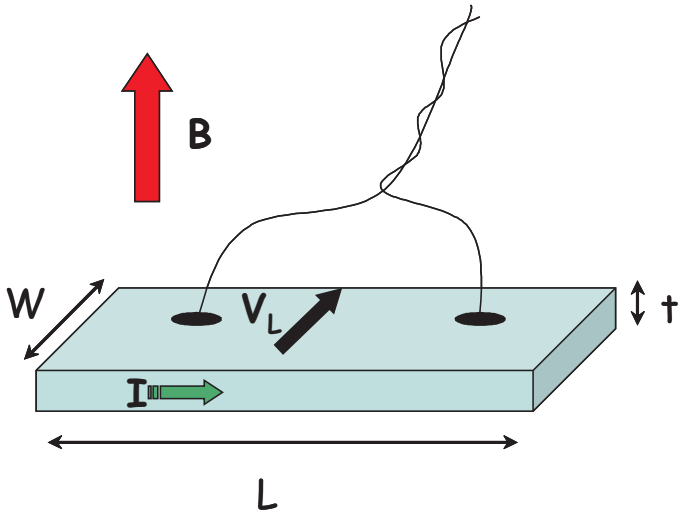}
\end{center}
\caption{Schematic view of the sample, showing the direction of the current $I$, the vortex velocity $V_L$ and the magnetic field $B$. $L$, $W$ and $t$ are length, width and thickness respectively.}
\label{fig1}
\end{figure}

\begin{figure}[!t]
\begin{center}
\includegraphics*[width=9.0cm]{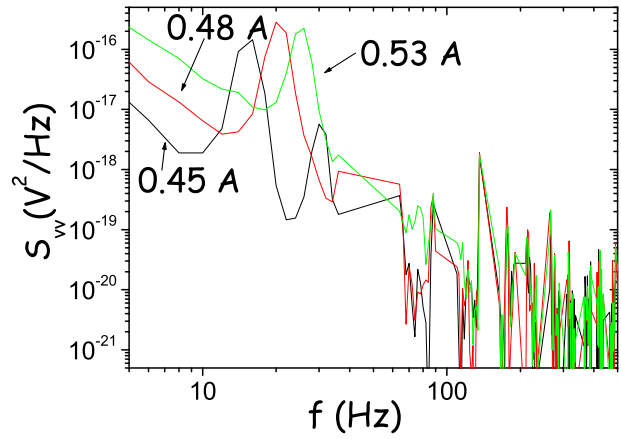}
\end{center}
\caption{Power spectrum of the voltage noise in the mixed state of Niobium (sample $\sharp 1$, B=$0.28$ T, T= $4.2$ K I= $0.45, 0.47$ and $0.53$ A), showing BBN and NBN at $f_c$ and at $2.f_c$ for $0.45$ A.}
\label{fig2}
\end{figure}

\begin{figure}[!t]
\begin{center}
\includegraphics*[width=18.0cm]{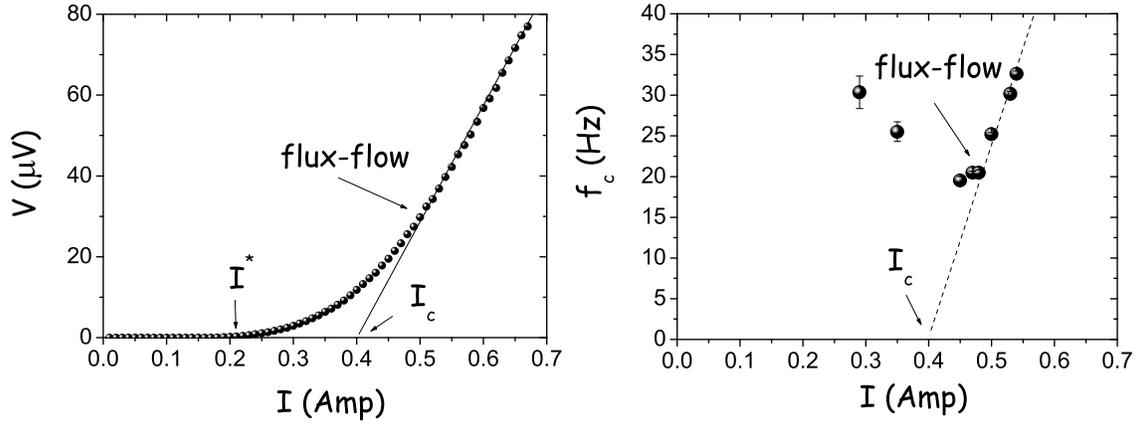}
\end{center}
\caption{Left: $V(I)$ curve of the sample $\sharp 1$, B=$0.28$ T, T= $4.2$ K. $I^*$ is the current where the first depinning occurs, $I_c$ is the average critical current of the whole sample.
 Right: The corresponding $f_c(I)$ curve. Note the linear relation between $f_c$ and $(I-I_c)$ as soon as the flux-flow is reached.}
\label{fig3}
\end{figure}

\begin{figure}[!t]
\begin{center}
\includegraphics*[width=9.0cm]{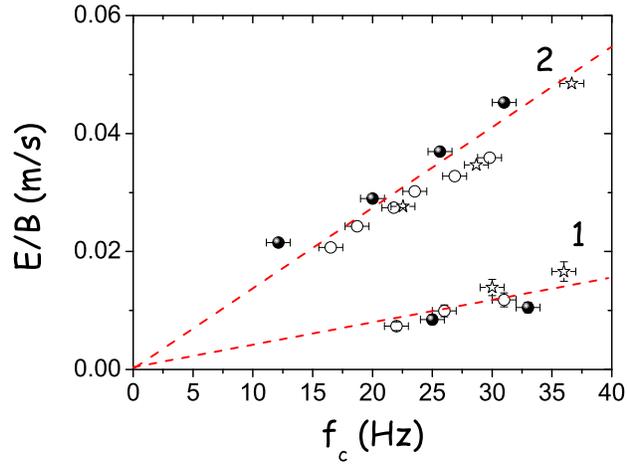}
\end{center}
\caption{$E/B$ as function of $f_c$ in the flux-flow regime: 1 is for the sample $\sharp 1$ (B=$0.24, 0.27, 0.28$ T below and in the peak effect region); 2 is for the sample $\sharp 2$ (B= $0.26T, 0.27, 0.28$ T, no peak effect).
 The dotted line is a linear fit which gives $1.30\pm 0.05 mm$ for the sample $\sharp 2$ and $0.35\pm 0.05 mm$ for the sample $\sharp 1$.}
\label{fig4}
\end{figure}

\begin{figure}[!t]
\begin{center}
\includegraphics*[width=9.0cm]{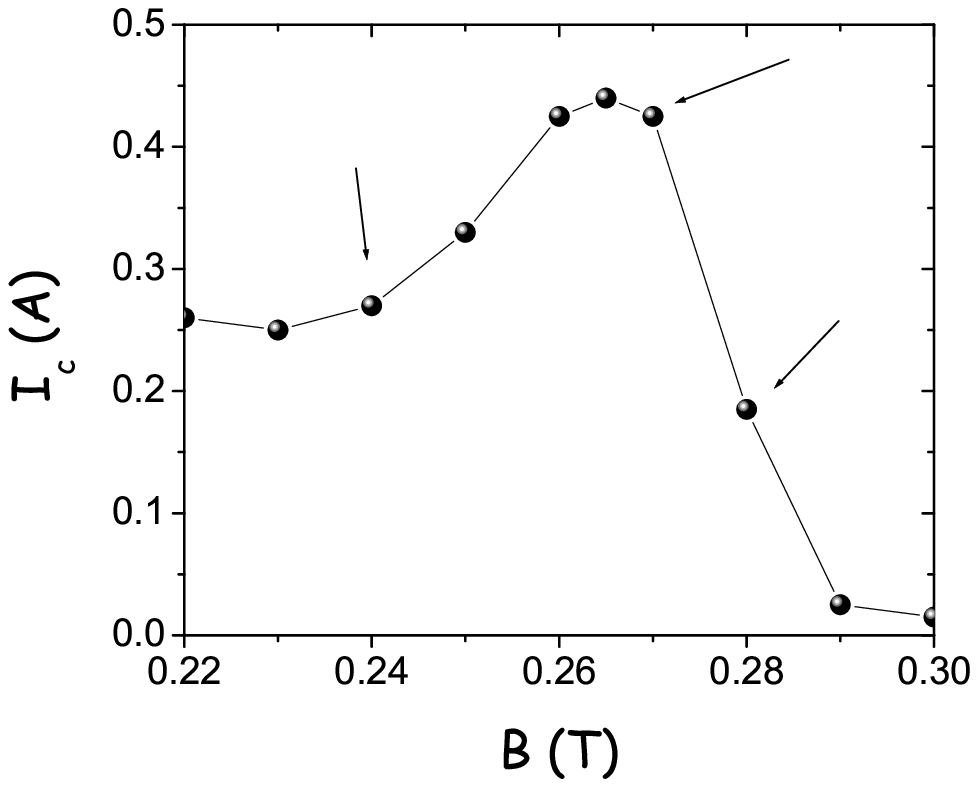}
\end{center}\caption{Critical current $I_{c}$ as function of the magnetic field for the sample $\sharp 1$. $I_{c}$ is defined in fig.3.
 Note that $I^*$, defined with a $10 nV$ criterium, follows the same variation. The arrows mark the field were the presence of NBN has been controlled.}
\label{fig5}
\end{figure}


\begin{references}
\label{sec:TeXbooks}
\bibitem{giam} T. Giamarchi , S. Bhattacharya, Vortex Phases in ``High Magnetic Fields: Applications in Condensed Matter Physics and Spectroscopy", C. Berthier et al., 9, 314,
 Springer-Verlag (2002).
\bibitem{kv} A.E. Koshelev and V. M. Vinokur, Phys. Rev. Lett. 73, 3580 (1994).
\bibitem{gld} P. Le Doussal and T. Giamarchi, Phys. Rev. B 57, 11356 (1998).
\bibitem{balents} L. Balents, M.C. Marchetti, and L. Radzihovsky, Phys. Rev. B 57, 7705 (1998).
\bibitem{zeldov} D.T. Fuchs, E. Zeldov, M. Rappaport, T. Tamegai, S. Ooi, and H. Shtrikman, Nature (London) 391, 373 (1998).
\bibitem{nous} A. Pautrat, C. Goupil, Ch. Simon, D. Charalambous, E. M. Forgan, G. Lazard, P. Mathieu, and A. Br$\hat{u}$let, Phys. Rev. Lett. 90, 087002 (2003).
\bibitem{nous2} A. Pautrat, J. Scola, Ch. Simon, P. Mathieu, A. Br$\hat{u}$let, C. Goupil, M. J. Higgins, and S. Bhattacharya,
Phys. Rev. B 71, 064517 (2005).
\bibitem{jones} R. G. Jones, E. H. Rhoderick, and A. C. Rose-Innes, Phys. Lett. 24A, 318 (1967).
\bibitem{olson} C.J. Olson, C. Reichhardt, and F. Nori, Phys. Rev. Lett. 81, 3757 (1998).
\bibitem{ryu} S. Ryu, M. Hellerqvist, S. Doniach, A. Kapitulnik, and D. Stroud, Phys. Rev. Lett. 77, 5114 (1996).
\bibitem{kolton2}  A.B. Kolton, D.Domínguez, and N. Gr$\oslash$nbech-Jensen, Phys. Rev. Lett. 83, 3061 (1999).
\bibitem{Enrik} E. Olive and J. C. Soret, Phys. Rev. B 77, 144514 (2008).
\bibitem{shobo}  M. Marchevsky, M. J. Higgins, and S. Bhattacharya, Phys. Rev. Lett. 88, 087002 (2002).
\bibitem{paltiel} Y. Paltiel, Y. Myasoedov, E. Zeldov, G. Jung, M. L. Rappaport, D. E. Feldman, M. J. Higgins, and S. Bhattacharya, 
Phys. Rev. B 66, 060503 (2002).
\bibitem{clem} J. R. Clem, Phys. Rep. 75, 1 (1981).
\bibitem{bernard2} B. Pla\c{c}ais, P. Mathieu, and Y. Simon, Phys. Rev. B 49, 15 813 (1994).
\bibitem{jojo} J. Scola, A. Pautrat, C. Goupil, and Ch. Simon, Phys. Rev. B 71, 104507 (2005). 
\bibitem{cdw} R. Fleming and C. C. Grimes, Phys. Rev. Lett. 42, 1423 (1979) .
              N. P. Ong, G. Verma and K. Maki, Phys. Rev. Lett. 52, 663 (1984). 
\bibitem{maeda}Y. Togawa, R. Abiru, K. Iwaya, H. Kitano and A. Maeda, Phys. Rev. Lett. 85 (2000).
\bibitem{danna} G. D'Anna, P. L. Gammel, H. Safar, G. B. Alers, D. J. Bishop, J. Giapintzakis, and D. M. Ginsberg, Phys. Rev. Lett. 75, 3521 (1995).
\bibitem{nbnheiden} C. Heiden, D. Kohake, W. Krings and L. Ratke, Journal of Low Temperature Physics 27, 1 (1977).
\bibitem{jo2} J. Scola, A. Pautrat, C. Goupil and Ch. Simon, Phys. Rev. B 73, 024508 (2006).
\bibitem{jo1} J. Scola, A. Pautrat, C. Goupil, L. M$\acute{e}$chin, V. Hardy, and Ch. Simon, Phys. Rev. B 72, 012507 (2005).
\bibitem{secondspectre} J.P. Restle, M.B. Weissman, G.A. Garfunkel, P. Pearah and H. Morkoc, Phys. Rev. B 34, 4419 (1986).
\bibitem{stm} A.M. Troyanovski, J. Aarts, P.H. Kes, Nature 399, 665 (1999).
\bibitem{thorel} Y. Simon and P. Thorel, Phys. Lett. 35, 450 (1971).
\bibitem{gougou} C. Goupil, A. Pautrat, Ch. Simon, P. G. Kealey, E. M. Forgan,
 S. L. Lee, S. T. Johnson, G. Lazard, B. Pla\c{c}ais, Y. Simon, P. Mathieu, R. Cubitt and Ch. Dewhurst, 
Physica C  341, 999 (2000). 
\bibitem{notebene} The flux-flow BBN in Niobium is about 1-2 orders of magnitude lower than in alloys for the same geometrical parameters,
 essentially because of the low normal state resistivity.
\bibitem{shobo2} A. C. Marley, M. J. Higgins, and S. Bhattacharya, Phys. Rev. Lett. 74, 3029 (1995).
\bibitem{mikeNb} R. D. Merithew, M. W. Rabin, M. B. Weissman, M. J. Higgins, and S. Bhattacharya, Phys. Rev. Lett. 77, 3197 (1996).

             
\end{references}
\end{document}